\documentclass[twocolumn,showpacs,preprintnumber,amsmath,amssymb,aps,prb]{revtex4}
\usepackage{graphicx}
\begin{document}

\title{Disordering Transitions and Peak Effect in Polydisperse Particle Systems}  
\author{C. Reichhardt and C.J. Olson Reichhardt} 
\affiliation{
Theoretical Division, 
Los Alamos National Laboratory, Los Alamos, New Mexico 87545}

\date{November 20, 2007}
\begin{abstract}
We show numerically that in a binary system of Yukawa particles,
a dispersity driven disordering transition 
occurs. In the presence of quenched disorder
this disordering transition coincides with a 
marked increase in the depinning threshold, 
known as a peak effect. 
We find that the addition of poorly pinned particles can 
increase the overall pinning in the sample by increasing the amount
of topological disorder present. If the quenched disorder is strong enough 
to create a significant amount of topological disorder in the monodisperse
system, addition of a poorly pinned species generates further disorder but
does not produce a peak in the
depinning force.
Our results indicate that for binary mixtures, optimal pinning occurs
for topological defect fraction densities of 0.2 to 0.25.
For defect densities below this range, the system retains orientational order.  
We determine the effect of the pinning density, strength, 
and radius on the depinning peak and
find that the peak effect is 
more pronounced in weakly pinning systems.   
\end{abstract}
\pacs{64.60.Cn,64.60.Ht,82.70.Dd,74.25.Qt}
\maketitle

\vskip2pc

\section{Introduction}

In two dimensional (2D) systems, an ordered 
phase can undergo an amorphization transition 
in the presence of random quenched disorder. 
At 
zero temperature and for sufficiently strong
quenched disorder, this transition 
is characterized by 
the appearance of isolated dislocations 
\cite{Fertig}. 
Disordering transitions can have a profound effect
on transport properties. 
One of the best known examples of this phenomenon 
is the peak effect observed for vortices 
in type-II superconductors \cite{Pippard,Kes1,Ling,Kes,Chaikin}.
Here, an ordered vortex lattice is weakly pinned by random quenched
disorder; however, as the temperature or magnetic field is increased,
the vortex lattice disorders and a sudden increase or peak in the
pinning force occurs.
The peak effect is known to be relevant
to 
2D and effectively 
2D superconducting systems \cite{Kes1}. 

An early explanation for the increase in the pinning force was 
that a disordered vortex system with numerous topological defects 
is much softer than an ordered vortex lattice, 
and as a result the vortices in the disordered state can shift
positions in order to accommodate to the pinning landscape 
\cite{Pippard,Chaikin}. 
A proliferation of topological defects 
has also been correlated with an increase in the effective
friction force in nanomaterials near the
bulk melting temperature where a 
peak effect type phenomenon can occur \cite{Timan,Holian}.  
Although the peak effect has been associated with the
appearance of topological defects, 
a comprehensive understanding of 
exactly how the density or type of topological 
defects relates to the pinning effectiveness is still lacking.
Interpreting the peak effect phenomenon is  
also complicated by thermal effects, and controversy remains over
whether the disordering of the vortex lattice is predominately a
melting phenomena \cite{Ling,Chaikin} or is nonthermal \cite{Forgan}. 
There are still several open issues related to 
the peak effect in superconductors
and its connection to
the shape of the current-voltage curves \cite{Zimanyi},  
transient dynamics \cite{Andrei}, and the 
magnetic field-temperature (H-T)
phase diagram \cite{Menon}.

In two dimensional systems with binary or polydisperse
interactions,   
a topological disordering transition occurs 
with increasing dispersity \cite{Stanley,Li,Onuki}. 
An open question is how quenched disorder affects a
dispersity-driven amorphization process 
and how the amorphization might alter the pinning effectiveness. 
A stiff lattice is poorly pinned by random
disorder, and thus in a monodisperse assembly 
of repulsively interacting particles 
on quenched random disorder, the depinning threshold decreases when
the lattice is stiffened by increasing the repulsion between the
particles \cite{Olson}. 
A similar decrease of the depinning threshold with increasing
particle-particle interactions occurs even when the monodisperse
system contains some topological disorder and is no longer a perfect
elastic lattice \cite{Olson}.
The situation may be different in a bidisperse system where the
relative strength of the repulsive interactions between the two particle
species can be independently tuned.
If the interaction strength of one species is increased while that of the
other species is held constant, 
the depinning threshold for motion over a random substrate may decrease;
however, an amorphization of the bidisperse lattice 
occurs when the difference between the repulsive interactions 
of the two species is large enough. 
The effective softness of the lattice increases sharply once topological
defects appear in the system, and thus
the depinning threshold may increase rather than decrease
when the particle interaction strength
is increased across the amorphization transition.

Polydisperse systems with quenched
disorder can be used as models for exploring the interaction between 
defects and pinning since the
number of topological defects can be controlled readily.
Binary and polydisperse particle interactions appear
in a variety of systems which can also have quenched disorder,
including vortices
in Bose-Einstein condensates \cite{Mueller}, 
electron bubble mixtures \cite{Kivelson},
mixtures of Abrikosov and Josephson vortices \cite{Bending}, 
and colloidal systems \cite{Maret,Bechinger}.
In two dimensional colloidal systems, it was recently demonstrated 
that interactions with a very long screening length can be 
realized experimentally \cite{Hsu} and that 
disordered phases occur which are likely induced by polydispersity \cite{Eric}.
Two dimensional colloidal disordering transitions and pinning phenomena 
in the presence of quenched disorder have 
also been demonstrated experimentally \cite{Ling2}.  
Interaction dispersity can also arise in the pinning of a nanofriction
system.
It would be very useful to understand how to control the defect density and  
pinning in these types of systems by 
manipulating the polydispersity in the particle interactions.  

In this paper we consider a two dimensional model of two species of 
particles interacting via a repulsive Yukawa potential 
both with and without quenched disorder. 
In the absence of quenched disorder, we find
a well defined low temperature disordering transition 
as a function of dispersity. In the
presence of weak quenched disorder, this disordering transition is accompanied 
by a sharp increase
in the depinning force or a peak effect. 
The peak effect phenomenon we
observe is nonthermal 
and occurs due to the proliferation of topological defects. 
Our results indicate that there is a complex interplay between the 
density of topological defects and the depinning force.  
We identify where the peak effect phenomenon 
occurs as a function of pinning strength, density, radius, and
interparticle interaction strength.

\section{Simulation}

We simulate a two dimensional system of bidisperse particles
in a sample of size $L_x=L_y=36$
with periodic boundary 
conditions in the $x$ and $y$ directions.
The particles interact via a
Yukawa potential 
\begin{equation} 
V(R_{ij}) =  
\frac{C_iC_jZ}{4\pi\epsilon\epsilon_0}\frac{e^{-\kappa R_{ij}}}{R_{ij}},
\end{equation}
where ${\bf R}_{i(j)}$ is the position of particle $i(j)$,
$R_{ij}=|{\bf R}_i-{\bf R}_j|$,
$C_{i(j)}$
is the charge of particle $i(j)$,
$1/\kappa$ is the screening length, and
$\epsilon$ is the dielectric constant. 
The interaction force prefactor $Z$ is set to $Z=1$ unless
otherwise noted.
In this work we fix kappa at $\kappa=2.0.$ 
We consider a binary arrangement of 
$N_A$ and $N_B$ particles of species $A$
and $B$, respectively,  which
have different 
charges denoted by 
$C_{A}$ and $C_{B}$. 
For a monodisperse system $C_{B}/C_{A} = 1.0$. 
The total number of particles $N = N_{A} + N_{B}$ and
the density $n=N/L_xL_y$.
In general we fix $N=856$ and $C_{A}=1$
and vary either the ratio $N_{B}/N$ or $C_{B}/C_A$. 
The particles also interact with a random quenched background which is modeled
as $N_p$ randomly placed parabolic traps of radius $r_{p} = 0.3$, density
$n_p=N_p/L_xL_y=0.66$ and maximum force $f_{p}$. 
The parabolic trap potential is similar to 
that used in previous work
\cite{Reichhardt},
$V_p(R_{ik})=-(f_p/2r_p)(R_{ik}-r_p)^2$ for $R_{ik}\le r_p$ and zero
interaction for $R_{ik}>r_p$, where $R_{ik}=|{\bf R}_i-{\bf R}_k|$ and
${\bf R}_k$ is the position of trap $k$.
The particles evolve under Brownian dynamics, performed by
integrating the overdamped equation of motion
\begin{equation}
\eta \frac{d{\bf R}_{i}}{dt} = -\sum^{N}_{j \ne i}\nabla V(R_{ij})  
- \sum^{N_{p}}_{k}\nabla V_{p}(R_{ik}) + {\bf F}_{D} + {\bf F}^{T},
\end{equation}
where $\eta$ is the damping constant. 
The external dc driving force ${\bf F}_{D}=F_D{\bf \hat{x}}$ 
is slowly increased from zero in increments of 
$\delta{\bf F}_D=2\times 10^{-5}$ 
applied every $2.5\times 10^4$ simulation steps. 
We have found that slower increment rates 
do not change the results. 
The thermal force ${\bf F}^{T}$ is modeled as random Langevin kicks
with $\langle {\bf F}^T\rangle=0$ 
and $\langle {\bf F}^T(t){\bf F}^T(t^\prime)\rangle=2\eta k_BT
\delta(t-t^\prime)$.  
Since we are interested in nonthermal effects, 
we consider a low temperature $T/T_m=0.15$, where $T_m$ is the melting
temperature of a monodisperse system 
with $N_B/N=0$ at density $n=0.66$.
The initial particle configurations are obtained using two techniques which
produce identical results.
In the first, we perform simulated annealing, while in the second, we place
the particles in a triangular lattice, and when the dispersity is large 
enough defects naturally appear. 

\begin{figure}
\includegraphics[width=3.5in]{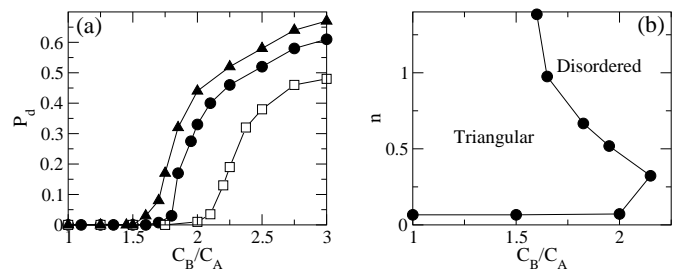}
\caption{
(a) Fraction of 
defected particles $P_d$ vs the ratio
of the particle
interaction strength $C_{B}/C_{A}$ for a mixture with $N_B/N=0.5$,
$C_{A} = 1.0$, and $f_p=0$ at different
net particle densities $n = 0.97$ (filled triangles), $n=0.66$ 
(filled circles), 
and $n=0.32$ (open squares). (b) 
Phase diagram of $n$ vs $C_{B}/C_{A}$ for 
the same system.
}
\end{figure}

\section{Effect of Particle Dispersity Without Quenched Disorder} 

To demonstrate that this system exhibits a dispersity 
driven disordering transition 
in the absence of quenched disorder, we consider a sample
with $N_B/N=0.5$ and $C_A=1.0$ at different particle densities $n=0.97$,
0.66, and 0.32 for $f_p=0$.
In Fig.~1(a) we plot the fraction of defected particles 
$P_d=N^{-1}\sum_{i=1}^N[1-\delta(6-z_i)]$
as a function of $C_{B}/C_{A}$ at the three different densities, 
where 
$z_i$ is the coordination number of particle $i$ obtained from
a Delaunay triangulation. 
In each case, when $C_B/C_A$ is near 1,
the system forms a triangular 
lattice free of topological defects and 
$P_d=0$. 
Once the dispersity is strong enough, a disordering transition 
occurs with a proliferation of defects and 
$P_d$ rises above zero.  
The dominant type of defects we observe are fivefold and sevenfold 
coordinated particles, although in the very disordered states 
it is possible to find a small fraction of fourfold or eightfold
coordinated particles.
As the density of the system increases, the dispersity
$C_B/C_A$ at which the disordering transition occurs decreases
since the particle-particle interactions are stronger at higher
density and thus the strain energy induced by the dispersity in the
particle interactions increases.

\begin{figure}
\includegraphics[width=3.5in]{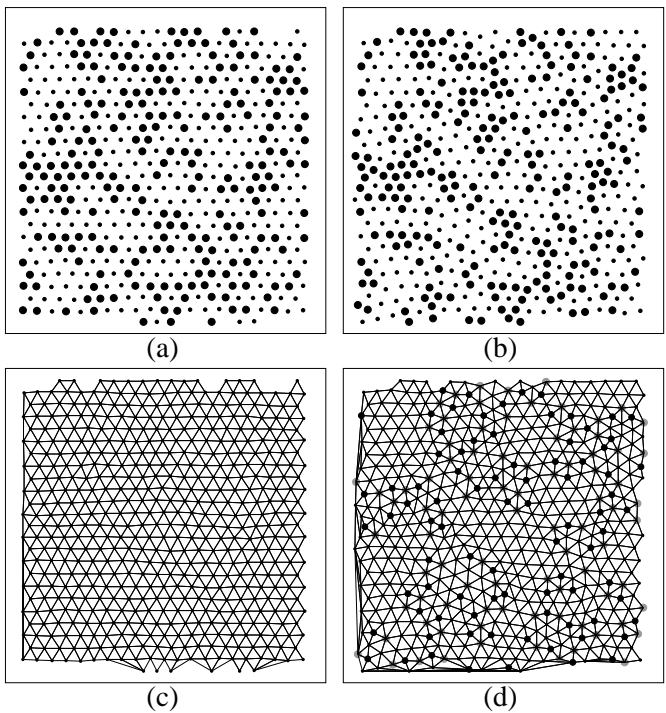}
\caption{(a,b) Particle positions in a 
$24 \times 24$ portion of a sample with $N_B/N=0.5$, $n=0.66$,
and $f_p=0$.
Large circles: species $A$; small circles: species $B$.
(a) The ordered regime at $C_B/C_A=1.2$. (b) 
The disordered regime at $C_B/C_A=2.0$.
(c,d) The corresponding Delaunay triangulations
at (c) $C_B/C_A=1.2$ and (d) $C_B/C_A=2.0$.  
Dark circles: fivefold coordinated particles; light circles:
sevenfold coordinated particles.
}
\end{figure}

By conducting a series of simulations
we map the order-disorder transition line
as a function of density $n$ and
polydispersity $C_B/C_A$, shown in Fig.~1(b). 
The order to disorder transition is 
defined as occurring when the fraction of defected particles reaches
$P_{d} = 0.2$. 
If a different cutoff value of $P_d$ is used to identify the transition, 
the general features of the phase diagram are unchanged but the 
precise location of the transition can shift.       
At low densities $n<0.06$ the system is in a
liquid state even for $C_B/C_A = 1.0$ since 
the particle-particle interactions are 
not strong enough to overcome the thermal forces.  
Since the simulation is performed at finite temperature, 
the thermal forces become important when the particle-particle 
interactions are weak enough. 
For a system with a finite particle-particle interaction range, 
the particle density can be decreased to a point at which there 
is no elasticity since adjacent particles are no longer interacting on 
average.  
This is what gives rise to the low density liquid state.  

In Fig.~2(a) we illustrate the particle positions for a system at $n=0.66$
with $C_B/C_A=1.2$ in the ordered regime.  The corresponding Delaunay 
triangulation shown in Fig.~2(c) indicates that all of the particles are
sixfold coordinated and form a triangular lattice.
Figure 2(b) shows the particle positions in the same system at 
$C_B/C_A=2.0$ in the disordered regime,  
while Fig.~2(d) illustrates that in this regime 
the lattice is filled with
fivefold and sevenfold coordinated defects.   

In the phase diagram of Fig.~1(b), we only distinguish between the 
topologically ordered and topologically disordered states, and indicate
the reentrant disordering transition into a liquid state that occurs at
low densities. 
At intermediate densities, the disordered phase may have glasslike features, 
implying that there could be another line on the phase diagram between a
liquidlike and glasslike state.  
Determining whether such a line is present or absent 
is beyond the scope of the present paper.
It is likely that the actual lowest energy state for the polydisperse system 
is phase separated; however, we have never observed such a state. 
Recent experiments and simulations of binary colloidal systems with 
dispersity similar to what we consider here
also produced no phase separated states, but did show
some evidence for clustering \cite{Maret}.  
In simulations by Sadr-Lahijany {\it et al.} \cite{Stanley}, 
an intermediate hexatic phase appeared 
in certain regions of the density-polydispersity phase diagram. 
The hexatic phase is characterized by an algebraic decay in the 
orientational correlation function. 
In general, hexatic phases are difficult to observe and there 
are still open questions about the nature of this phase. 
It is beyond the scope of this work to address the 
possible existence of a hexatic phase in clean systems; 
instead, we focus on the regimes with quenched disorder.

\begin{figure}
\includegraphics[width=3.5in]{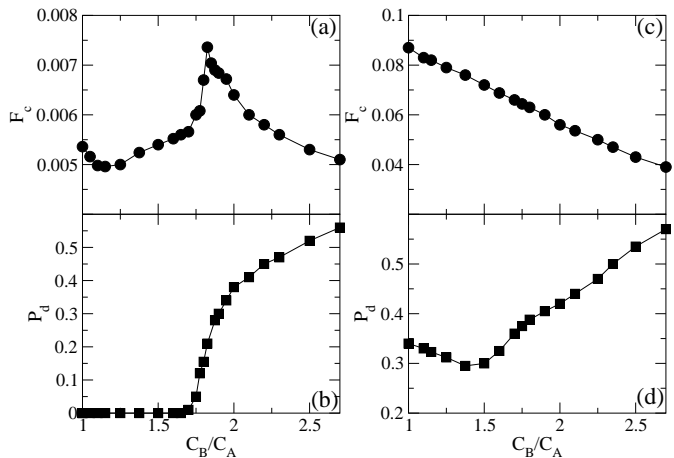}
\caption{
(a) Depinning force $F_{c}$ and (b) fraction of 
defected particles $P_{d}$ vs $C_{B}/C_{A}$ for a
mixture at $n=0.66$ with $N_B/N=0.5$ on
a random pinning substrate with $f_{p} = 0.1$ and $n_{p} = 0.66$.
(c,d) $F_c$ and $P_d$ versus $C_B/C_A$ 
for a system with the same parameters except with $f_{p} = 0.4$. 
}
\end{figure}

\section{Effect of Quenched Disorder} 

In Fig.~3(b) we plot $P_{d}$ vs $C_{B}/C_{A}$ for a 
mixture with $N_B/N=0.5$ and $n=0.66$ in the presence of
a random pinning potential with  $f_{p} = 0.1$ and $n_p=0.66$. 
At this pinning strength the 
system is free of dislocations for $C_{B}/C_{A} < 1.6$. 
Near $C_{B}/C_{A} = 1.75$ an order-disorder transition occurs. 
In Fig.~3(a) we plot the
corresponding critical depinning force $F_{c}$ which is the value of $F_D$ 
at which
the velocity of the particles in the direction of the drive exceeds 
$5\times 10^{-4}$. In general there are two predominant pinning regimes. 
For the defect-free lattice at $C_B/C_A<1.6$, 
the depinning occurs elastically without the further generation of 
defects. In this case the depinning threshold
is well defined as the entire lattice moves at the same average velocity. 
Once defects begin to appear around $C_B/C_A=1.75$, the
depinning becomes plastic and a portion of the particles 
can be moving while another portion remains pinned, leading to 
very inhomogeneous and intermittent velocity bursts. 
The large fluctuations associated with plastic depinning can also 
lead to long transient responses where a portion of the system 
is initially moving when a drive is applied, but over time the entire 
system becomes pinned.
In order to accurately measure a depinning threshold in the 
topologically disordered regime, sufficiently long run times at each
force increment are needed in order to obtain a smoother velocity-force curve. 
We also note that changing the value of the threshold velocity 
for the depinning measurement does not change the general features of 
the results.
  
Figure 3(a) shows that the depinning force $F_c$
initially decreases with increasing
$C_{B}/C_{A}$ in the ordered regime, but then starts to shift upward once
$C_B/C_A=1.2$. At the order-disorder transition 
which occurs at $C_B/C_A=1.75$ there is 
a sharp increase in $F_{c}$ followed by a 
slow decrease for increasing $C_{B}/C_{A}$. We
term this phenomenon
a {\it dispersity driven peak effect}. 
This result demonstrates that a peak effect can 
arise via a completely nonthermal disordering process.

For a monodisperse system with $C_B/C_A=1$, increasing 
the magnitude of the repulsive interaction 
$C_B$
always reduces 
$F_{c}$ \cite{Olson}. 
In contrast, the results in Fig.~3(a) show that
increasing the charge of only a 
fraction of the particles can actually 
{\it increase} the depinning force. 
The maximum in $F_{c}$ occurs when a fraction 
$P_d=0.2$ of the
particles are defected. 
As $C_{B}/C_{A}$ increases further above $C_B/C_A=1.75$, 
$P_d$ increases;
however, the depinning force decreases. 
This result indicates that although a 
peak effect occurs at the onset of defect proliferation,
simply having more topological defects does not directly
translate to a higher depinning force.

For a monodisperse
system, increasing the 
strength of the repulsive particle-particle interactions 
decreases the depinning threshold. 
In the bidisperse system in Fig.~3(a), when $C_{B}/C_A$ is increased 
above 1 by only a small amount, 
the net distortion of the lattice is small and no dislocations
are induced since the system is still nearly monodisperse. 
Thus in this regime, increasing $C_{B}/C_A$ increases the overall 
effective lattice stiffness and causes the depinning threshold
to decrease.
Once $C_{B}/C_A$ is large enough [$C_B/C_A>1.2$ in Fig.~3(a)], 
more significant distortions 
of the lattice occur due to the fact that the two particle species
are randomly interspersed, and these distortions effectively soften the
lattice and increase the depinning threshold.
As $C_B/C_A$ continues to increase, the distortions become large enough to
induce the formation of dislocations [$C_B/C_A=1.75$ in Fig.~3(a)], 
and the order-disorder transition occurs. 
The depinning threshold increases rapidly since the particles can shift
into optimal pinning locations once the dislocations appear,
and the maximum in $F_{c}$ occurs at $C_B/C_A=1.8$. 
When $C_{B}/C_A$ is increased above 1.8, defects continue to proliferate in
the lattice but the particles are already in their optimal pinning locations
so no further enhancement of the depinning threshold occurs.
Instead, as the particle-particle interaction force 
becomes stronger, the local stiffness of the lattice increases,
shifting the particles away from the optimal pinning locations 
and causing the depinning threshold to decrease again with
increasing $C_B/C_A>1.8$.      
  
To show that the order-disorder transition is responsible for the 
peak in the depinning force, in Fig.~3(c,d) 
we plot $F_{c}$ and $P_{d}$ vs $C_{B}/C_{A}$ for a system with
the same parameters as in Fig.~3(a,b) 
but with a stronger pinning force of $f_p=0.4$. 
For this value of $f_{p}$, the monodisperse system $C_{B}/C_{A} = 1.0$ already
contains a significant fraction of topological defects, 
with $P_d=0.34$. 
As $C_{B}/C_{A}$ increases from 1 the 
defect density $P_d$ decreases;
however, for 
$C_{B}/C_{A} > 1.5$ 
$P_d$ begins to increase again. 
In contrast, the depinning force $F_{c}$ decreases 
monotonically for all $C_{B}/C_{A} > 1.0$. 
We have performed a series of simulations for 
other values of $f_{p}$ and pinning densities $n_p$ and
find the following general features. 
(1) A peak effect phenomenon occurs whenever there
is an order to disorder transition. 
(2) The peak in the depinning force occurs when a fraction of about 
$P_d=0.15$ to $0.3$ of the particles are defected. 
(3) If the system is already strongly disordered for the monodisperse 
case $C_B/C_A=1$, there is no enhancement in 
depinning force with increasing $C_B/C_A$ even 
when the fraction of topological defects increases.

\begin{figure}
\includegraphics[width=3.5in]{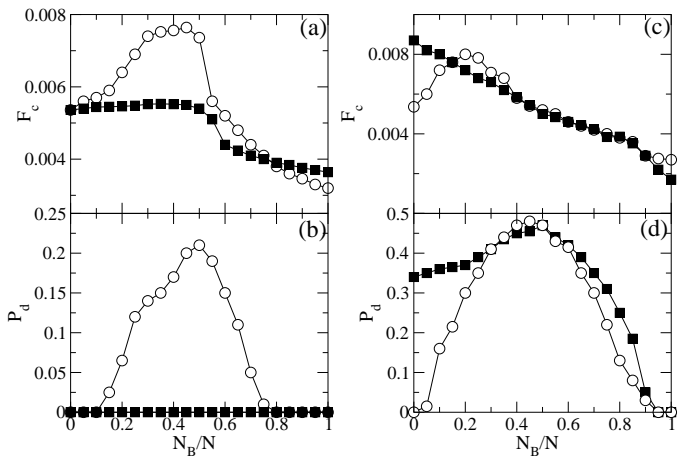}
\caption{
(a) The depinning force $F_{c}$ 
for varied species ratio $N_{B}/N$ at fixed $n = 0.66$, $n_{p} = 0.66$,
and  $f_{p} = 0.1$. 
Filled squares: $C_B/C_A=1.5$. 
Open circles: $C_{B}/C_{A} = 1.825$. 
(b) The corresponding
fraction of defected particles $P_{d}$. 
(c) 
$F_{c}$ vs $N_{B}/N$ for 
the same set of parameters in (a)
but with $C_{B}/C_{A} = 2.5$.  Open circles: $f_p=0.1$.
Filled squares: 
$f_{p} = 0.4$ (this curve was divided by 10 for presentation purposes). 
(d) The corresponding $P_{d}$ curves.   
}
\end{figure}
       
\section{Changing the Particle Species Ratio} 

We next consider the effect of varying the species ratio 
$N_{B}/N$  
at fixed particle density $n=0.66$ to explore the disordering
and depinning effects 
as the mixture varies from monodisperse species $A$, $N_B/N=0$, to
monodisperse species $B$, $N_B/N=1$.
In Fig.~4(a) we plot the depinning force $F_c$ 
vs $N_{B}/N$ for a system with $f_{p} = 0.1$, 
$n_{p} = 0.66$, and $C_{A} = 1$
at $C_B/C_A=1.5$ and $C_B/C_A=1.825$.
Figure 4(b) shows the corresponding $P_{d}$ curves.
For $C_{B}/C_A = 1.5$,
there is no dislocation induced transition at any fraction $N_{B}/N$ and 
$F_{c}$ shows no enhancement but merely decreases above $N_B/N=0.5$.   
The depinning force $F_c=0.003$ for the monodisperse $B$ system at $N_B/N=1$
is less than that of the pure $A$ system,
$F_c=0.005$ at $N_B/N=0$, as expected 
for monodisperse systems due to the increased repulsive force between 
species $B$ particles compared to species $A$.
The depinning force for the pure $B$ system is lower at
$C_B/C_A=1.825$ than at $C_B/C_A=1.5$; however, for 
intermediate values of $N_B/N$ in the $C_B/C_A=1.825$ sample
there is a 
strong enhancement of $F_{c}$ over the depinning force for either the
pure $A$ or pure $B$ systems.
The enhancement of
$F_{c}$ for $C_{B}/C_A = 1.825$ 
is associated with the creation of defects, as shown by the
behavior of $P_{d}$ in Fig.~4(b). 
The defect density reaches a maximum value 
of $P_d=0.21$ at $N_{B}/N = 0.5$. 
The peak value of $F_{c}$ also falls
at $N_B/N=0.5$. 
These results
show that, due to induced disorder, the effective pinning force for 
mixed species can be higher than that of either of the pure
species.
     
In Fig.~4(c) we plot $F_{c}$ vs  $N_{B}/N$ for a system with $C_{B}/C_A = 2.5$
and $C_{A} = 1.0$ at $f_{p} = 0.1$ and $f_{p} = 0.4$. 
For $f_{p} = 0.1$, the system becomes increasingly disordered
as $N_B/N$ increases from zero until $P_d$ reaches a maximum 
of $P_d=0.48$ at $N_B/N=0.44$, as shown in Fig.~4(d).  For $N_B/N>0.44$,
$P_d$ decreases back to $P_{d} = 0$ for the pure $B$ system.
There is a peak in $F_{c}$ at $N_{B}/N = 0.2$, 
where $P_{d} = 0.3$. 
In general we find that for 
increasing $C_{B}/C_A$ and 
fixed $C_{A}$, the peak in $F_{c}$
shifts to lower values of $N_{B}/N$. 
The peak in $F_{c}$ occurs 
when a fraction $P_d\sim 0.2$
of the particles are disordered, 
and since higher values of $C_{B}/C_A$ disorder the system more effectively,
a lower fraction of species B is needed 
to reach $P_d=0.2$ as $C_B/C_A$ increases. 
This result suggests that
strong enhancement of the pinning can be 
achieved by adding a few strongly
repulsive particles to a pure system.   
For the case of $f_{p} = 0.4$ in Fig.~4(c,d), the 
pure $A$ case is already strongly defected, so as $N_B/N$ is increased from
zero, $F_{c}$ monotonically
decreases while $P_{d}$ increases slightly and then decreases. 
 
We find that for decreasing $f_{p}$, the maximum value 
$F_c^{max}$ of $F_c$ in the mixed
sample increases relative to the value 
$F_c^{pure}$ of $F_c$ in the pure samples with $N_B/N=0$ or 1.
This can be understood by
a simple argument. 
For the pure samples, the pinning is collective
and $F_{c}^{pure} \propto F^{2}_{p}$, 
while at the mixed sample peak the pinning behaves more like single 
particle pinning
with $F_{c}^{max} \approx F_{p}$, 
so that $F^{max}_{c}/F^{pure}_{c} \propto 1/F_{p}$.
This indicates that in pure systems with weak pinning, the addition of 
a second species 
could have a very significant effect on the pinning properties.  

\begin{figure}
\includegraphics[width=3.5in]{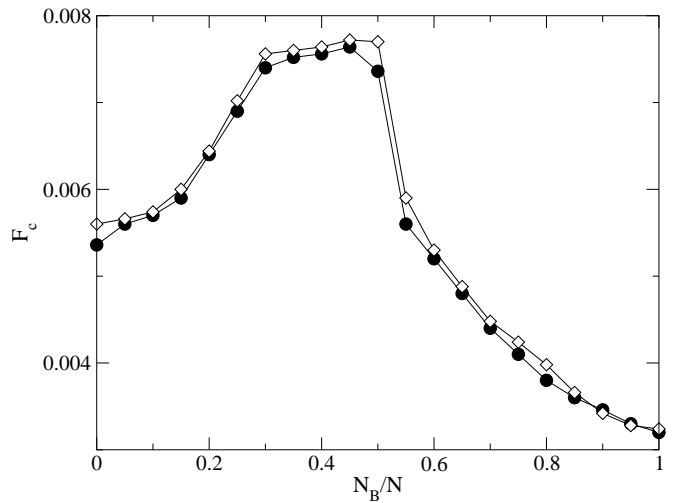}
\caption{
(a) The depinning force $F_{c}$ 
for varied species ratio $N_{B}/N$ at fixed $n = 0.66$, $n_{p} = 0.66$,
$f_{p} = 0.1$, 
and  $C_B/C_A=1.5$. Open diamonds: $T/T_{m} = 0.$ 
Filled circles: $T/T_{m} = 0.15.$     
}
\end{figure}

The results up to this point 
were obtained at a finite but low temperature. 
Most experiments with colloids are performed
in regimes where some Brownian motion occurs, and our results 
show that the peak effect phenomenon is robust against the addition of a
small temperature and thus could be observed in colloidal experiments.
At higher temperatures, a significant amount of creep occurs in the 
disordered regimes and makes obtaining an accurate depinning 
threshold difficult. This type of thermally activated motion
will be described elsewhere. 
To illustrate that our results remain unchanged for $T = 0$, in Fig.~5 
we plot $F_{c}$ vs $N_{B}/N$ for the same system in Fig.~4(a) 
at $C_{B}/C_{A} = 1.5$, showing that there is a negligible difference
between the $T=0$ and finite $T$ results.

\section{Varied Pinning Density and Pinning Radius}

\begin{figure}
\includegraphics[width=3.5in]{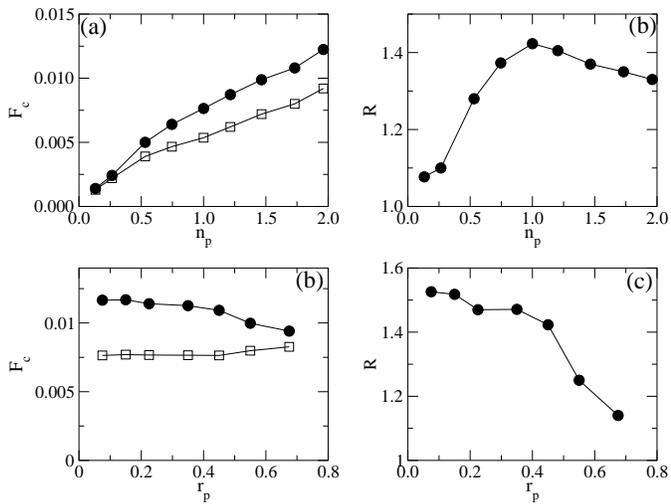}
\caption{
(a) The depinning force $F_{c}$ vs pinning density $n_{p}$ for a system with 
fixed $f_{p} = 0.1$, $r_{p} = 0.3$, and $N_B/N=0.5$ for 
the bidisperse system $C_{B}/C_{A}  = 1.85$ (black circles) 
and the monodisperse system $C_{B}/C_{A} = 1.0$ (open squares).
(b) The ratio $R$ of $F_c$ for the bidisperse and monodisperse
samples in (a).  
(c) The depinning force $F_{c}$ vs 
pinning radius $r_{p}$ for a fixed pinning density of $n_{p} = 0.66$
at $N_B/N=0.5$ for $C_{B}/C_{A} = 1.85$ (black circles) and 
$C_{B}/C_{A} = 1.0$ (squares).   
(d) The ratio $R$ of $F_{c}$ for the bidisperse and monodisperse
samples in (c). 
}
\end{figure}

We next vary several parameters of the pinning sites 
in order to understand how general the 
peak effect phenomenon is in binary systems with quenched disorder.
In Fig.~6(a) we show $F_{c}$ vs pinning density $n_{p}$ for a 
sample with $N_B/N=0.5$, $n=0.66$, $n_p=0.66$, and $f_p=0.1$ at 
$C_{B}/C_{A} = 1.825$, corresponding to the peak value of $F_c$
in Fig.~3(a), and $C_B/C_A=1.0$, corresponding to a monodisperse system.
To indicate the magnitude of the peak effect, we calculate the ratio
of the bidisperse and monodisperse depinning thresholds,
$R=F_{c}(C_B/C_A=1.825)/F_{c}(C_B/C_A=1)$,
and plot the result in Fig.~6(b).
Although $F_{c}$ increases monotonically with $n_{p}$ 
for both values of $C_{B}/C_{A}$, $F_c$ is always higher in
the polydisperse system.
The ratio $R$ goes through a maximum 
near $n_{p} = 0.66$ when the density of particles equals the density
of pins.
At low $n_{p}$, the bidispersity leads to little enhancement of
the depinning threshold compared to the monodisperse system.
When there are few pinning sites, most pinning sites can be occupied 
even in the monodisperse case, so softening the lattice by making the
particles bidisperse and introducing dislocations 
does not create a significantly larger number
of pinned particles.  As a result, the enhancement of $F_c$ by the
polydispersity is weak in this regime.
Similarly, at high pinning densities there are so many pins available
that less distortion of the lattice is required to permit most of 
the particles to occupy pinning sites, so the introduction 
of dislocations due to bidispersity does not create as large of an 
enhancement in $F_c$ compared to the intermediate pinning density regime.       

In Fig.~6(c) we plot $F_{c}$ versus pinning radius 
$r_{p}$ for a system with fixed $n_{p} = 0.66$, $n=0.66$,
$N_B/N=0.5$ and $f_{p} = 0.1$ at $C_{B}/C_{A} = 1.825$ and 
$C_{B}/C_{A} = 1.0$. 
We plot the corresponding ratio $R$ of the depinning forces in
Fig.~6(d).  Here, $R$ decreases
monotonically with increasing $r_{p}$. 
As the size of the pinning sites increases, 
the amount of lattice distortion required to permit most particles to
occupy a pinning site decreases since a larger area of the sample is
covered by the pinning sites.
Thus, there is a decreasing enhancement of $F_c$ due to bidispersity as
$r_p$ increases.

\section{Varied Pinning Strength and Particle Interaction Strength}  

\begin{figure}
\includegraphics[width=3.5in]{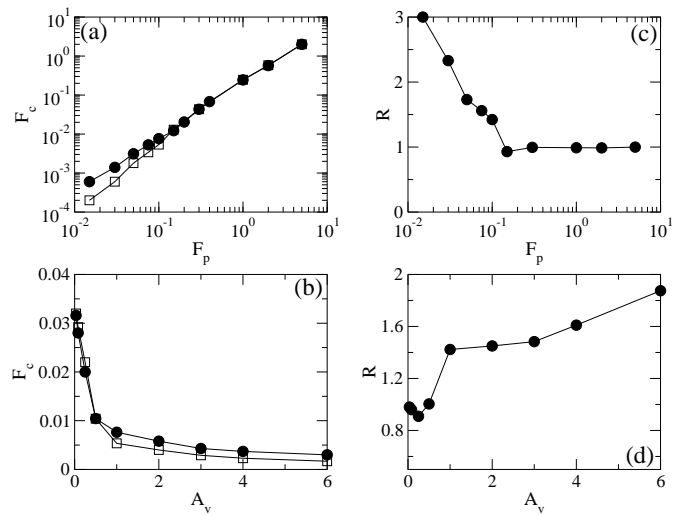}
\caption{
(a) The depinning force $F_{c}$ vs 
pinning strength $f_{p}$ for a system with fixed $n_{p} = 0.66$, 
$n=0.66$, $r_{p} = 0.3$, and $N_B/N=0.5$. 
Filled circles: $C_{B}/C_{A} = 1.825$; open squares: $C_{B}/C_{A} = 1.0$. 
(b) The ratio $R$ of the 
bidisperse and monodisperse critical depinning forces from (a).     
(c) The depinning force $F_{c}$ vs 
particle interaction strength $Z$ for a system with fixed $n_{p} = 0.66$,
$n=0.66$, $r_{p} = 0.3$, and $N_B/N=0.5$. 
Filled circles: $C_{B}/C_{A} = 1.825$; 
open squares: $C_{B}/C_{A} = 1.0$.
(d) The ratio $R$ of the critical depinning forces from (c). 
}
\end{figure}

In Fig.~7(a) we plot $F_{c}$ vs 
pinning strength $f_{p}$ for a system with $n_{p} = 0.66$, 
$n=0.66$, $r_{p} = 0.3$, and $N_B/N=0.5$ for the bidisperse case
$C_{B}/C_{A} = 1.825$ and the monodisperse case $C_{B}/C_{A} = 1.0$. 
The corresponding ratio $R$ of the bidisperse to the monodisperse depinning
threshold is plotted in 
Fig.~7(b).
For large $f_{p}$, both the monodisperse 
and the bidisperse systems are considerably defected and 
$R\approx 1$. In this regime the critical
depinning force $F_c$ depends linearly on $f_{p}$, 
as expected for single particle pinning behavior. For
$f_{p} < 0.2$, the 
monodisperse system is defect-free and the 
depinning crosses over to collective pinning behavior
with $F_{c} \propto f_{p}^2$. 
In the same regime, true collective behavior does not occur for 
the bidisperse system and
the depinning force falls off less steeply than $f_{p}^2$, 
as seen in Fig.~7(a). 
Thus the ratio $R$ between the 
bidisperse and monodisperse depinning thresholds grows with 
decreasing $f_{p}$, as shown in Fig.~7(b) for $f_{p}<0.2$.
This result indicates that for weaker pinning, 
the peak effect phenomenon is more pronounced. 

To vary the overall particle-particle interaction strength, we adjust the
value of the interaction prefactor $Z$ in Eq. (1).
For larger $Z$, the particles are more repulsive and the lattice is stiffer.    
We plot the effect of changing $Z$ on $F_c$ in Fig.~7(c) for
both monodisperse ($C_B/C_A=1$) and bidisperse 
($C_B/C_A=1.825$) samples with   
$n_{p} = 0.66$, $n=0.66$, $r_{p} = 0.3$, $f_{p} = 0.1$, 
and $N_B/N=0.5$.
The corresponding ratio $R$ of the depinning threshold for the bidisperse and
monodisperse samples is shown in
Fig.~7(d).
We note that changing $Z$ has an effect similar to
changing $\kappa$ since the lattice becomes stiffer for smaller $\kappa$.
For $Z > 1$, the 
monodisperse system at $C_B/C_A=1$ is defect free, the depinning 
is elastic, and $R$ increases with increasing $Z$. 
In this case, the effect of increasing
$Z$ is similar to the effect of decreasing $f_{p}$. 
For $Z < 1$, the monodisperse system becomes topologically 
disordered and $R$ approaches $1.0$ as $Z$ decreases.  

These results show that the peak effect phenomenon is observable
in a polydisperse system whenever the effective quenched disorder is weak 
enough that the monodisperse system is free of defects.
The monodisperse lattice is pinned collectively by the random disorder, and
the collective pinning is always weaker than the single particle pinning which
occurs in a topologically disordered polydisperse particle lattice.
In regimes where the quenched disorder is strong enough that 
both the monodisperse and the polydisperse systems
are topologically disordered, there is little difference in the 
depinning force between the polydisperse and the monodisperse cases, 
and no peak in the depinning force appears.    

\section{Defect Density and Orientational Order} 

\begin{figure}
\includegraphics[width=3.5in]{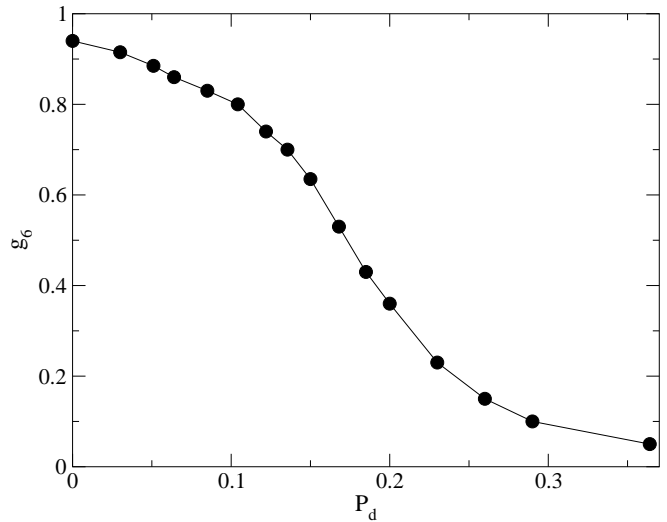}
\caption{
The value of the bond orientational correlation function $g_6(r)$ at
$r = 5.0$ versus the fraction of defected particles $P_{d}$ in 
a system with $C_{B}/C_{A} = 2.5$ and $f_p=0.1$ at various particle ratios
$N_B/N$.
}
\end{figure}

A general trend visible in
Fig.~3(a,b) and Fig.~4(a,b,c,d) is that the peak in $F_{c}$ corresponds to 
a defect density of around $P_d \approx 0.2$. For defect densities $P_d>0.2$,
the depinning force $F_c$ decreases with increasing $P_d$.
Recent simulations of two-dimensional monodisperse 
Yukawa particle systems have shown that as a function of defect density,
the orientational correlations are lost when the 
fraction of defected particles is 
$P_d > 0.2$ \cite{Qi}. The orientational
correlations are measured by means of the 
local bond orientational order parameter
\cite{Strandburg},
\begin{equation}
\psi_{6}({\bf R}_{i}) = \left\langle\frac{1}{N_i}\sum^{N_i}_{j =1}e^{6i\theta_{ij}} \right\rangle .
\end{equation}
Here $N_i$ is the number of nearest neighbors of particle $i$ and 
$\theta_{ij}$ is the angle between an arbitrary fixed reference
axis and the bond connecting particles $i$ and
$j$. 
The bond orientational correlation function 
$g_{6}(r) = \langle\psi_{6}^*({\bf R}^\prime)\psi_6({\bf R}^\prime - {\bf r})
\rangle/g(r)$, where the pair distribution function 
$g(r)=\langle\delta({\bf R}^\prime)\delta({\bf R}^\prime-{\bf r})\rangle.$ 
Orientational order is present when $g_6(r)$ decays algebraically with $r$.

We measure 
$g_6(r)$
as a function of defect density $P_d$ in 
a system with $C_{B}/C_{A} = 2.5$ and $f_{p} = 0.1$ 
over a range of defect fractions from $P_d=0.0$ to $P_d=0.5$ 
obtained by varying $N_B/N$, 
as indicated in Fig.~4(d). 
In Fig.~8 we plot the value of $g_{6}(r)$ at $r = 5.0$ versus $P_{d}$. 
The orientational order decreases monotonically with increasing 
$P_{d}$.
We also find that $g_6(r)$
decays exponentially with $r$ for $P_{d} > 0.2$, indicating that the
system has lost orientational order above $P_d=0.2$.

For defect densities below $P_d=0.2$, the system is orientationally 
ordered and the depinning is 
elastic in nature. The defects soften the lattice locally, 
allowing the particles to become better pinned. 
As more defects are added and $P_d$ increases, the lattice continues
to soften until the orientational
order is lost at $P_d=0.2$, which corresponds to the elastic-plastic 
depinning transition.
Once in the plastic depinning regime, the addition of more 
particles or more defects 
does not further increase the softness of the lattice.
In our system, species $B$ is more highly charged than species $A$ and thus
species $B$ is in general less well pinned due to the higher 
strength of the particle-particle interaction compared to the particle-pin
interaction.
As $N_B/N$ increases from zero, the defect density of the entire system
increases, which would normally lead to more effective pinning, but
clumps of species $B$ particles can form
and create localized poorly pinned areas, which may decrease the overall
pinning effectiveness.
Although it is beyond the scope of this paper to address,    
an open question is to understand exactly why a fraction of 
$P_d=0.2$ defects corresponds
to the loss of orientational order in two-dimensional systems 
and whether this fraction $P_d=0.2$ of defects is 
universal to all two-dimensional systems 
at the orientational disordering transition
regardless of the form of the particle-particle interactions.

\section{Summary} 
    
To summarize, we have shown that for a binary assembly of
repulsively interacting Yukawa particles, 
there can be an order to disorder dispersity driven transition 
induced by the addition of a more strongly repulsive species.
In the presence of
weak quenched disorder, the dispersity driven transition 
coincides with a peak effect in the
depinning force. 
The depinning force for a mixture of particles can be higher 
than that for either of the pure species. 
If the quenched disorder is strong and the pure species sample is
already disordered, the addition of a more strongly repulsive species 
monotonically decreases the depinning force. 
For a completely ordered system, if the addition of a new species 
does not induce
dislocations, there is no enhancement of the depinning force.   
We also find that the optimal pinning occurs when the fraction of
non-sixfold coordinated particles is around $0.2$, which correlates with
the loss of orientational order.
Our results suggest that a peak effect phenomenon 
can occur without thermal fluctuations and
that the effectiveness of the pinning can be tuned 
by adjusting the particle mixture.     

This work was carried out under the auspices of the NNSA of the U.S. DoE
at LANL under Contract No. DE-AC52-06NA25396.


\begin{thebibliography}{99}

\bibitem{Fertig}
M.-C. Cha and H.A.~Fertig, 
Phys.~Rev.~Lett.~{\bf 74}, 4867 (1995). 

\bibitem{Pippard}
A.B.~Pippard, Philos.~Mag.~{\bf 19}, 217 (1969).

\bibitem{Kes1} 
P.H.~Kes and C.C.~Tsuei, Phys.~Rev.~B {\bf 28}, 5126 (1983);
A.C.~Marley, M.J.~Higgins, and S.~Bhattacharya, Phys.~Rev.~Lett.~{\bf 74},
3029 (1995). 

\bibitem{Ling}
X.S.~Ling,
S.R. Park, B.A. McClain, S.M. Choi, D.C. Dender, and J.W. Lynn,
Phys.~Rev.~Lett.~{\bf 86}, 712 (2001). 

\bibitem{Kes}
A.M.~Troyanovski, 
M. van Hecke, N. Saha, J. Aarts, and P.H. Kes,
Phys.~Rev.~Lett.~{\bf 89}, 147006 (2002).   

\bibitem{Chaikin}
C.~Tang, X.S.~Ling, S.~Bhattacharya, and P.M.~Chaikin, 
Europhys.~Lett.~{\bf 35}, 597 (1996). 

\bibitem{Timan}
T.~Zykova-Timan, D.~Ceresoli, and E.~Tosatti, 
Nature Mater. {\bf 6}, 230 (2007). 

\bibitem{Holian}
J.E.~Hammerberg, B.L.~Holian, T.C.~Germann, and R.~Ravelo,
Metall.~Mater.~Trans.~A {\bf 35}, 2741 (2004).

\bibitem{Forgan}
E.M.~Forgan, 
S.J. Levett, P.G. Kealey, R. Cubitt, C.D. Dewhurst, and D. Fort,
Phys.~Rev.~Lett.~{\bf 88}, 167003 (2002).

\bibitem{Zimanyi}
S.~Bhattacharya and M.J. Higgins, Phys.~Rev.~Lett.~{\bf 70}, 2617 (1993),
M.J.~Higgins and S.~Bhattacharya, Physica C {\bf 257}, 232 (1996); 
C.J.~Olson, G.T.~Zim{\' a}nyi, A.B.~Kolton, and 
N.~Gr{\o}nbech-Jensen, Phys.~Rev.~Lett.~{\bf 85},
5416 (2000).    

\bibitem{Andrei}
W.~Henderson, E.Y.~Andrei, M.J.~Higgins and S.~Bhattacharya, 
Phys.~Rev.~Lett.~{\bf 77}, 2077 (1996).   

\bibitem{Menon}
G.I.~Menon, Phys.~Rev.~B {\bf 65}, 104527 (2002). 

\bibitem{Stanley}
M.R. Sadr-Lahijany, P.~Ray, and H.E.~Stanley,
Phys.~Rev. Lett. {\bf 79}, 3206 (1997).  

\bibitem{Li}
M.~Li, Phys.~Rev.~B.~{\bf 62}, 13979 (2000). 

\bibitem{Onuki}
L.~Foret and A.~Onuki,
Phys.~Rev.~E {\bf 74}, 031709 (2006).

\bibitem{Olson}
C.J. Olson, C.~Reichhardt, and S.~Bhattacharya, Phys. Rev. B {\bf 64},
024518 (2001).  

\bibitem{Mueller}
E.J.~Mueller and T.L.~Ho, Phys.~Rev.~Lett.~{\bf 88}, 180403 (2002);
K.~Kasamatsu, M.~Tsubota, and M.~Ueda, Phys. Rev. Lett. {\bf 91},
150406 (2003). 

\bibitem{Kivelson}
E.~Fradkin and S.A.~Kivelson, Phys.~Rev.~B {\bf 59}, 8065 (1999). 

\bibitem{Bending}
A.~Grigorenko,
S. Bending, T. Tamegai, S. Ooi, and M. Henini,
Nature {\bf 414}, 728 (2001). 

\bibitem{Maret}
N.~Hoffmann, F.~Ebert, C.N.~Likos, H.~L\" owen, and G.~Maret,
Phys.~Rev.~Lett.~{\bf 97}, 078301 (2006).  

\bibitem{Bechinger}
J.~Baumgartl, R.P.A. Dullens, M. Dijkstra, R. Roth, and C. Bechinger,
Phys.~Rev.~Lett. {\bf 98}, 198303 (2007).

\bibitem{Hsu}
M.F.~Hsu, E.R.~Dufresne, and D.A.~Weitz, Langmuir {\bf 21}, 4881 (2005).  

\bibitem{Eric}
E.R.~Dufresne {\it et al.}, to be published.   

\bibitem{Ling2}
A.~Pertsinidis and X.S.~Ling, Phys.~Rev.~Lett., in press (2007).

\bibitem{Reichhardt}
C. Reichhardt and C.J. Olson, Phys.~Rev.~Lett.~{\bf 89}, 078301 (2002);
C. Reichhardt and C.J. Olson Reichhardt, Phys.~Rev.~E {\bf 75}, 040402(R) 
(2007). 

\bibitem{Qi}
WK Qi, Y. Chen, and SM Qin, arXiv:0709.2035.

\bibitem{Strandburg}
K.J. Strandburg, Rev. Mod. Phys. {\bf 60}, 161 (1988).

\end{thebibliography}
\end{document}